\documentclass[twocolumn,showpacs,preprintnumbers,amsmath,floatfix]{revtex4}
\usepackage{graphicx}
\usepackage{dcolumn}
\usepackage{bm}

\begin{document}

%
\title{Energy and Angular Distribution of Upward \ UHE Neutrinos and Signals of Low Scale Gravity: Role of Tau Decay}
\author{Shahid Hussain}
 \email{vacuum@ku.edu}
\author{Douglas W. McKay}%
 \email{mckay@kuark.phsx.ku.edu}
\affiliation{Department of Physics \& Astronomy 
University of Kansas, Lawrence, KS 66045.}

\date{\today}

\begin{abstract}
We present extensive results and analysis of energy and angular distributions
of diffuse UHE $\nu _{e}$, $\nu _{\mu }$, and $\nu _{\tau }$ 
fluxes propagated through earth, with and without augmentation of the 
standard model interactions by low scale gravity. With propagated fluxes in hand we 
estimate event rates in a $1km^{3}$ detector in ice
with characteristics of ICECUBE. We determine that, at 0.5PeV energy threshold, 
there is a 
significant difference in the ratios of down shower events to upward muon events 
between the standard model and the low scale gravity cases with 1TeV and 2TeV mass scales. 
The same is true for energy 
threshold at 5PeV. Though the difference is large in all flux models, statistical
significance of this difference depends on the flux models, especially at 5PeV and above. 
Both flavor 
assumptions, $\nu _{e}$, $\nu _{\mu }$, $\nu _{\tau }::1$, $2$, $0$ and $%
\nu _{e}$, $\nu _{\mu }$, $\nu _{\tau }::1$, $1$, $1$, and all flux models show large
differences. Though rates of tagged events are low, we find that $\nu _{\tau
}$ regeneration by $\tau $ decay may play an important role in disclosing
deviations from standard model predictions at energies in the neighborhood
of 1 PeV for 1TeV-scale gravity, for example. We emphasize those analyses
whose sensitivity to new physics is independent of the flux model assumed.
\end{abstract}

\maketitle

\section{Introduction}

The pursuit of high and ultra-high energy neutrinos has greatly intensified
over the past decade as more and more neutrino telescopes have entered the
search. Though the observation of MeV-neutrinos emitted from SN 1987a is
over 15 years old \cite{kam}, \cite{imb}, there is still no firm candidate
for TeV, PeV or EeV neutrinos of galactic or extra-galactic origin. Yet
there is good reason to expect a neutrino flux exists in this energy regime
because of the great success of air-shower detectors in building a detailed
record of cosmic rays with these very-high to ultra-high energies \cite
{crayrev}. The photons or nuclear particles that are generally believed to
initiate the observed shower are accompanied by neutrinos with similar
energies, in most models of the high-energy particle emission by the
sources. In any case, the neutrinos emitted by production and decay of pions
by the highest energy primary cosmic rays as they interact with the cosmic
microwave background, the so-called GZK \cite{gzk} neutrinos, should be
present at some level at ultra-high energies \cite{gzknu}, regardless of the mechanism
responsible for producing the observed cosmic rays \cite{fly,agasa,hires}.
Even if there are no super-GZK neutrinos, there are a number of models \cite
{sdss,wb,rjproth,manb,sozabo,engel,sigl,wichos}, which predict the existence
of neutrinos in the PeV-EeV range. By choosing several contrasting flux
models and using enhanced cross sections from low scale gravity \cite{add,rs,aadd}
, we look for new physics effects that are relatively independent of flux
models.

The expanding experimental capabilities and the strong theoretical interest
in understanding the physics of astrophysical sources and particle
interactions of the highest energy cosmic rays makes it imperative to study
all aspects of the neutrino observation process. The number of groups
reporting limits on fluxes and projecting improved limits with expanded data
sets or with new facilities is impressive. In the range 1 TeV to 1 PeV, the
AMANDA \cite{aman}, Frejus \cite{frejus}, MACRO \cite{macro} and Baikal \cite
{baik} experiments have reported limits on neutrinos from astrophysical
(non-atmospheric) sources. In the range 1 PeV to 1 EeV, AGASA \cite{agasa2},
AMANDA \cite{hundertmark}, Fly's Eye \cite{fly2}, and RICE \cite{limits}
have all reported limits. Above 1 EeV, AGASA, Fly's Eye, GLUE \cite{glue}
and RICE all put limits on the flux that extend up into the GZK range. The
upper limits are getting interestingly close to the predictions of several
models and actually below the predictions in several cases. The situation is
heating up and will get hotter as the experiments like AUGER, which is
already reporting preliminary results on air showers \cite{auger} and
ICECUBE \cite{icecubelim} are fully operational. Meanwhile, expanded data
sets and improvements in sensitivity in experiments like RICE will continue
to search and to push down on limits until the first UHE neutrinos are
observed \cite{webdetect}.

These detection capabilities that have been achieved and will be improved
and expanded in the next few years have direct impact on particle physics.
The detection estimates, upon which limits are based, all rely on the
extrapolation of neutrino cross sections well beyond the currently measured
energy range. Is QCD correctly predicting these cross sections \cite
{qcdcross}? Is there new physics that enhances neutrino cross sections at
high energies \cite{lsg}? What is the effect of new neutrino interactions 
\cite{prop2,highest,cosmicBH,ring1,ring2,haim,renouhe} or neutrino mixing 
\cite{renouhe,mix,dbbeac} on the expected rates of detection in various
telescopes? Clearly there is ample motivation for examining the consequences
of various combinations of assumptions about the physics governing the cross
sections and the assumptions about the flavor composition of the
astrophysical flux of neutrinos. What, if any, are the observable
distinctions among the various possibilities of flux and interaction
characteristics? These questions and the experimental prospects for answers
motivate this work.

There is considerable published work on $\tau $-neutrino propagation through
earth in \textit{standard model (}SM\textit{)} using analytic and
computational tools \cite
{smphalzen,smpnaumov,smpbeacom,smpreno,smpbecattini,smphettlage,fargionf,
smpfeng,smpweiler}
in the scenarios $\nu _{e}$, $\nu _{\mu }$, $\nu _{\tau }::1$, $2$, $0$ and $%
\nu _{e}$, $\nu _{\mu }$, $\nu _{\tau }::1$, $1$, $1$, and some analytical
and computational work on neutrino propagation in low scale gravity (LSG)
models has also been done in the $\nu _{e}$, $\nu _{\mu }$, $\nu _{\tau }::1$%
, $2$, $0$ scenario \cite{lsgneuprop1, prop2}. A detailed study has not been
done in the $\nu _{e}$, $\nu _{\mu }$, $\nu _{\tau }::1$, $1$, $1$ scenario
in LSG models. In this paper we solve, using Runge-Kutta method \cite{numrec}%
, the coupled differential equations for the four leptons $\nu _{e}$, $\nu
_{\mu }$, $\nu _{\tau }$, and $\tau $ in both of the above scenarios, in SM
and LSG models. For cross section calculations, we use Gaussian and monte
carlo integration methods \cite{numrec} with CTEQ6-DIS parton distributions 
\cite{cteq6}. Our results confirm significant regeneration effect due to
taus in the SM as already shown by several authors \cite{smpreno,
smpbecattini}. However, as we will see the regeneration due to taus is not
as significant in LSG models. Also, by comparing results of \cite{smpreno,
smpbecattini}, one finds that electromagnetic (EM) losses of $\tau $ are not
making a significant difference in the SM\ fluxes of $\nu _{\tau }$ around
1PeV, hence, we do not include EM losses in our work here. As we will see in
the next section, EM\ losses are not important at all in LSG\ models.

In Section 2 we talk about cross sections and interaction lengths in SM and
LSG ; Section 3 gives the formalism for neutrino propagation through the
earth; in Section 4 we show our results for different neutrino models and
discuss them; in Section 5 we develop formalism for event rates calculation
and in Section 6 we show and discuss our results for event rates; Section 7
gives the summary of our results and the conclusion.

\section{Cross Sections and Interaction Lengths}

Before evaluating the equations of propagation for the four leptons $\nu
_{e} $, $\nu _{\mu }$, $\nu _{\tau }$, and $\tau $ through the earth in next
section, we need to calculate their cross sections on isoscalar nucleons ($N=%
\frac{p+n}{2}$), where n stands for neutron and p for proton. We need to
calculate their neutral current (NC) and charged current (CC) weak
interaction cross sections in SM and eikonal (EK) and black hole (BH) cross
sections in LSG models. The LSG models do not discriminate among different
particles; they are the same for all the four leptons $\nu _{e}$, $\nu _{\mu
}$, $\nu _{\tau }$, and $\tau $. The SM total cross sections for different
flavors are also the same within a few percent at the ultra high energies we
are interested in here. We will assume the SM cross sections are the same
for all the four leptons $\nu _{e}$, $\nu _{\mu }$, $\nu _{\tau }$, and $%
\tau ,$ and we will use $\nu _{\mu }$ NC and CC\ cross sections for all of
them. The differences between $\tau $ and $\nu $ weak cross sections are not
important because $\tau $ decay is the only dominant process for $\tau
^{\prime }s$ up to energies $10^{8}GeV,$ and at higher energies the
differences are not significant anyway \cite{tauinter}. A detailed
discussion of SM $\nu _{\mu }-nucleon$ cross sections is given in Ref.\cite
{gandhi}. We have used CTEQ6-DIS parton distributions. Details of
calculation for LSG model cross sections are given in Ref.. \cite{jmpr,prop2}%
, where CTEQ4-DIS parton distributions \cite{cteq4} were used. The
difference between CTEQ4-DIS and CTEQ6-DIS cross section calculations is not
significant. We outline the LSG calculation here so that the presentation is
reasonably self contained.

The classical gravity Schwarzschild radius $r_{S}(\surd s)$ is the dominant
physical scale when the collision energy is large compared to the Plank
mass. At impact parameters smaller than $r_{S}$, we use the parton-level
geometrical cross section \cite{bh} 
\begin{equation}
\hat{\sigma}_{\mathrm{BH}}\approx \pi r_{S}^{2}.  \label{sigma_BH}
\end{equation}
For values of the classical impact parameter, b, larger than $r_{s}$ we use
the contributions to the amplitude in the eikonal approximation. In Eq. \ref
{sigma_BH}, $r_{S}$ is the Schwarzchild radius of a $4+n$ dimensional black
hole of mass $M_{\mathrm{BH}}=\sqrt{\hat{s}}$ \cite{mp}, 
\begin{equation}
r_{S}={\frac{1}{M}}\left[ \frac{M_{\mathrm{BH}}}{M}\right] ^{\frac{1}{1+n}%
}\left[ {\frac{2^{n}\pi ^{(n-3)/2}\Gamma \left( {\frac{3+n}{2}}\right) }{2+n}%
}\right] ^{\frac{1}{1+n}},
\end{equation}
where $\sqrt{\hat{s}\text{ }}$is the neutrino- parton C.M. energy, and $M$
is the 4+n-dimensional scale of quantum gravity. Multiplying by the parton
distribution functions, $f_{i}(x,q)$, choosing q at a value characteristic
of black hole production and integrating over momentum fraction x, gives the
estimate 
\begin{equation}
\sigma _{\nu N\rightarrow BH}(s)=\sum_{i}\int_{x_{min}}^{1}dx\hat{\sigma}%
_{BH}(xs)f_{i}(x,q).
\end{equation}
We take $x_{min}=M^{2}/s$ and $q=\surd \hat{s}$. The dependence of $\sigma
_{\nu N\rightarrow BH}(s)$ on the choice of $x_{min}$ and the treatment of q
is discussed in Ref. \cite{AFGS,cavagahn,yashnam,idaoda}.

For the input amplitude to the eikonal approximation, referred to as the
Born amplitude, we choose 
\begin{equation}
i\mathcal{M}_{\mathrm{Born}}=\sum_{j}{\frac{ics^{2}}{M^{2}}}{\frac{1}{%
q^{2}+m_{j}^{2}},}
\end{equation}
where $c$ is the gravitational coupling strength, $c=(M/\overline{M}%
_{P})^{2} $ and $\overline{M}_{P}=2.4\times 10^{18}$GeV is the reduced, four
dimensional Planck mass. Here $q=\sqrt{-t}$ is the usual lepton momentum
transfer. The index \textit{j} must include the mass degeneracy for the 
\textit{j}th K-K mode mass value. The sum, which can be well approximated by
an integral, must be cut off at some scale, generally taken to be of the
order of M. The transverse Fourier transform of the Born amplitude produces
the eikonal phase as a function of impact parameter b, 
\begin{equation}
\chi (s,b)={\frac{i}{2s}}\int {\frac{d^{2}q}{4\pi ^{2}}}\exp (i\mathbf{q}%
\cdot \mathbf{b})i\mathcal{M}_{\mathrm{Born}}.
\end{equation}
Evaluating the integral over q and representing the sum in the Born term by
an integral, one finds the ultraviolet-finite result 
\begin{eqnarray}
\chi (s,b) &=&-{\frac{s(2^{2n-3}\pi ^{{\frac{3n}{2}}-1})}{M^{n+2}\Gamma (n/2)%
}}2\int_{0}^{\infty }dmm^{n-1}K_{0}(mb)  \nonumber \\
&=&\left( {\frac{b_{c}}{b}}\right) ^{n},
\end{eqnarray}
where 
\begin{equation}
b_{c}^{n}={\frac{1}{2}}(4\pi )^{{\frac{n}{2}}-1}\Gamma \left[ {\frac{n}{2}}%
\right] {\frac{s}{M^{2+n}}}\ .
\end{equation}
The eikonal amplitude is then given in terms of the eikonal phase by 
\begin{eqnarray}
\mathcal{M} &=&-2is\int d^{2}b\exp (i\mathbf{q}\cdot \mathbf{b})\left[ \exp
(i\chi )-1\right]  \nonumber \\
&=&-i4\pi s\int dbbJ_{0}(qb)\left[ \exp (i\chi )-1\right] .
\end{eqnarray}
The eikonal amplitude can be obtained analytically \cite
{Emparan1,Emparan2,grw2} in the strong coupling $qb_{c}>>1$ and weak
coupling $qb_{c}<<1$ limits.

For strong coupling, the stationary phase approximation is valid, yielding 
\begin{equation}
\mathcal{M}=A_{n}e^{i\phi _{n}}\left[ {\frac{s}{qM}}\right] ^{\frac{n+2}{n+1}%
}\ ,
\end{equation}
where 
\begin{equation}
A_{n}={\frac{(4\pi )^{\frac{3n}{2(n+1)}}}{\sqrt{n+1}}}\left[ \Gamma \left( {%
\frac{n}{2}}+1\right) \right] ^{\frac{1}{1+n}}\ ,
\end{equation}
\begin{equation}
\phi _{n}={\frac{\pi }{2}}+(n+1)\left[ {\frac{b_{c}}{b_{s}}}\right] ^{n},
\end{equation}
and $b_{s}=b_{c}(qb_{c}/n)^{-1/(n+1)}$. In the cross section calculation, we
set the amplitude equal to its value at $q=1/b_{c}$ for values of q that are
less than $1/b_{c}$, since the small q region makes negligible contribution
to the cross section.

We assume that the black hole cross section is the dominant one for $q\geq
1/r_{S}$. The eikonal cross section is cut off at this value of q, since it
is not expected to be reliable for values of q larger than $1/r_{S}$ in any
case. 
\begin{figure}
\includegraphics[width=2.2in,angle=-90]{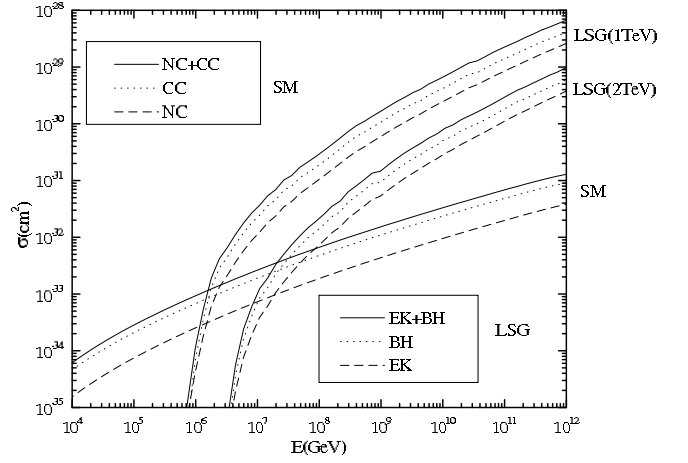}
\caption{\label{fig:f1}$\nu _{\mu }$--isoscalar nucleon
cross sections ($cm^{2}$) vs energy ($GeV$): for low scale gravity (LSG)
models, with number of extra dimensions n=6, we plot eikonal (EK, dashed
line), black hole (BH, dotted line), and total (EK+BH, solid line) cross
sections; for standard model (SM) we plot neutral current (NC, dashed line),
charged current (CC, dotted line), and total (NC+CC, solid line) cross
sections.}
\end{figure}
\begin{figure}
\includegraphics[width=2.4in,angle=-90]{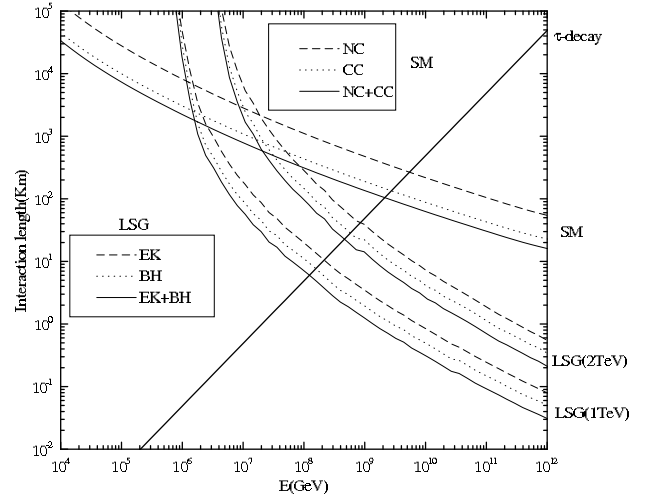}
\caption{\label{fig:f2}Interaction lengths ($km$) vs
energy ($GeV$): interaction length $\mathcal{L}_{int}=\frac{1}{\sigma
(E)N_{A}\rho }$, where $\sigma (E)$ are plotted in Fig. 1, $N_{A}$ is
Avogadro's number, and $\rho $ is the material density; we choose $\rho
=8g.cm^{-3}$ to make some comparisons with Ref. \protect\cite{smpbeacom} .
For low scale gravity (LSG) models, with number of extra dimensions n=6, we
plot eikonal (EK, dashed line), black hole (BH, dotted line), and total
(EK+BH, solid line) interaction lengths. For standard model (SM) we plot
neutral current (NC, dashed line), charged current (CC, dotted line), and
total (NC+CC, solid line) interaction lengths. We also plot $\tau $-decay 
length (thick solid line).}
\end{figure}

In Fig.1. we plot the SM and LSG (models with mass scale $1TeV$ and $2TeV;$ $%
n=6$) neutrino-isoscalar nucleon cross sections. For SM we plot neutral
current (NC), charged current (CC), and total ( NC + CC) cross sections. For
LSG we plot eikonal ($EK$), black hole ($BH$), and total (EK+BH) cross
sections. We see $\sigma _{BH}$ is larger then $\sigma _{EK}$ for our case ($%
n=6$). However, in $n=3$ case not shown here, the reverse is true. Our
results turned out to be only marginally sensitive to the number of
dimensions, so we choose to work with $n=6$, for which the bound on the
scale M is the weakest.

The interaction length in a material with density $\rho $ is defined here as 
$\mathcal{L}_{int}=(\frac{1}{N_{A}\rho \sigma })$, where $N_{A}$ is
Avogadro's number and $\sigma $ is the cross section for the interaction.
Fig.2. gives the interaction lengths in SM and LSG as well the $\tau $-decay
length. We set $\rho =8.0gcm^{-3}$ to make some comparisons between the
interaction lengths and the electromagnetic (EM) ranges of taus and muons
given in the Ref.\cite{smpbeacom}, using this value of $\rho $.

Which particles do we need to include in our propagation of neutrinos
through the earth? We have six candidates which might be coupled with each
other: $\nu _{e}$, $\nu _{\mu }$, $\nu _{\tau }$, and their three leptonic
partners. One can exclude electrons from this list, both in SM and LSG,
because they shower electromagnetically before they produce any $\nu _{e}$
through the CC interaction at energies of our interest here. The case for $%
\mu ^{\prime }s$ and $\tau ^{\prime }s$ needs some attention because it is
different in SM and LSG. In SM we can ignore $\mu ^{\prime }s$ but not $\tau
^{\prime }s$ because of the comparatively smaller decay length and larger EM
ranges of taus. For LSG, muons play some role in the propagation through the
earth at high enough energy as explained below. The effect is quite small,
however.

If we look at the interaction lengths and $\tau $-decay length in Fig.2, we
find that: (i) the LSG1 model (LSG model with mass scale 1TeV) interaction
lengths become smaller than the $\tau $-decay length for $E>10^{8}GeV$. This
implies the regeneration effect due to taus will be suppressed in LSG models
(ii) if we look at the EM ranges of taus \cite{smpbeacom}, we find that the
LSG1 model interaction lengths become smaller than the tau EM range for $%
E>10^{8}GeV$. This gives a reason for not including, in LSG, the EM energy
losses of taus in our propagation of neutrinos, coupled with taus, through
the earth. (iii) If we look at the EM ranges of muons \cite{smpbeacom}, we
find that even the muon EM energy losses are not important for $E>10^{8}GeV.$
(iv) Given the above reasons, interestingly, taus and muons become almost
identical in LSG for $E>10^{8}GeV.$ This means one may have to treat muons
and taus on equal footing in the propagation of neutrinos through the earth
for $E>10^{8}GeV$ in LSG1 model and for $E>10^{9}GeV$ in LSG2 model (LSG
model with mass scale 2TeV). However, in the present work, we do not include
muons in the propagation equations because we are looking at neutrinos
around 1PeV here, and we expect the coupling of muons with the $\nu _{\mu }$
via CC interaction for $E>10^{8}GeV$ will not affect the neutrino flux much
around 1PeV by feed down; the muon decay length, being so large in contrast
to that of $\tau ,$ will play no role in regeneration of $\nu _{\mu }$ \cite
{smpbeacom}.

\section{ Neutrino Propagation Through Earth}

Here we discuss the coupled propagation of $\nu _{e}$, $\nu _{\mu }$, $\nu
_{\tau }$, and $\tau $ through earth. We do not include the EM energy losses
of taus for the reasons discussed in the previous section. Suppose we have a
differential flux\footnote{%
All of our fluxes include neutrinos and antineutrinos.} $F^{i}(E,x,\theta )$
of lepton of species i at the surface of earth, then the transport equation
for each of the four leptons is 
\begin{widetext}
\begin{eqnarray}
\frac{dF^{i}(E,x,\theta )}{dx} &=&-\text{ }N_{A}\rho (x,\theta
)F^{i}(E,x,\theta )\sigma _{t}^{i}(E)-\frac{F^{i}(E,x,\theta )}{\mathcal{L}%
_{dec}^{i}(E)}+\sum_{j}[N_{A}\rho (x,\theta )\int_{E}^{\infty }dE^{^{\prime
}}F^{j}(E^{^{\prime }},x,\theta )\frac{d\sigma ^{j\rightarrow
i}(E^{^{\prime }},E)}{dE}  \nonumber \\
&&+\int_{E}^{\infty }dE^{^{\prime }}F^{j}(E^{^{\prime }},x,\theta )%
\frac{dP^{dec(j\rightarrow i)}(E^{^{\prime }},E)}{dE}],  \label{propeq}
\end{eqnarray}
\end{widetext}
\begin{figure}
\includegraphics[width=1.0in]{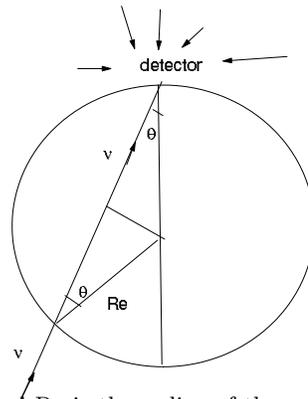}
\caption{Drawing of the earth: $\theta $ is the nadir angle and Re is the
radius of the earth. Arrows outside the earth represent the down flux.
Upflux is the flux coming through the earth.}
\label{fig:f3}
\end{figure}
where the first two terms give the loss and the last two terms give gain of
the flux per unit length in the same energy bin E. Avogadro's number N$_{A}$
times the density $\rho (x,\theta )$, gives the number of target nucleons
per unit volume at the nadir angle $\theta $ and distance x in the earth
(see Fig. 3.). We use the earth density model from Ref. \cite{earthdens}. $%
\sigma _{t}^{i}(E)$ is the total cross section for a lepton of flavor i to
interact with a nucleon and be expelled from the energy bin E:

\[
\sigma _{t}^{i}(E)=\sigma _{CC}^{i}(E)+\sigma _{NC}^{i}(E)+\sigma
_{BH}^{i}(E)+\sigma _{EK}^{i}(E),
\]
where we use the same $\sigma _{t}^{i}(E)$ for all the four leptons $\nu _{e}
$, $\nu _{\mu }$, $\nu _{\tau }$, and $\tau $ (LSG cross sections are the
same for them; see Section 2 for discussion on SM cross sections.). The
second term in Eq.\ref{propeq} gives the loss due to decays. It is zero for
the neutrinos, and for taus 
\begin{equation}
\mathcal{L}_{dec}^{\tau }(E)=\gamma c\mathcal{T},  \label{ltdec}
\end{equation}
where $\gamma =\frac{E_{\tau }}{m_{\tau }}$ is the Lorentz factor, $\mathcal{%
T}$ is the mean life time of taus and c is the speed of light in vacuum. The
third term in Eq.\ref{propeq} gives the gain in the flux of species $i$ in
the bin $E$, resulting from interaction of the species $j$ at $E^{^{\prime
}}>E$, and 
\begin{eqnarray}
\sum_{j}\frac{d\sigma ^{j\rightarrow i}(E^{^{\prime }},E)}{dE} &=&\frac{%
d\sigma _{NC}^{i\rightarrow i}(E^{^{\prime }},E)}{dE}+\frac{d\sigma
_{CC}^{j\rightarrow i(i\neq j)}(E^{^{\prime }},E)}{dE}  \nonumber \\
&&+\frac{d\sigma _{EK}^{i\rightarrow i}(E^{^{\prime }},E)}{dE}.  \nonumber
\end{eqnarray}
In the above equation, there is no need for the second term on the right
hand side for $\nu _{e}$ and $\nu _{\mu }$ propagation equations as we are
not keeping track of electrons and muons for the reasons given in the last
section. However, we keep this term for the $\nu _{\tau }$ and $\tau $
equations as we are propagating taus along with the neutrinos. For the
reasons given earlier, we use the same NC, CC, EK differential cross
sections for all the four leptons in the above equation. The fourth term in
Eq.\ref{propeq} is the gain in flux of species i in the bin E due to tau
decays. This term is zero in the tau flux equation. For $\nu _{e}$, $\nu
_{\mu }$, and $\nu _{\tau }$ we consider the corresponding decay channels of
taus as formulated in \cite[2000]{smpreno} :(i) $\tau \rightarrow \nu _{\tau
}\mu \nu_{\mu }$, (ii) $\tau \rightarrow \nu _{\tau }e\nu _{e}$, (iii) 
$\tau \rightarrow \nu _{\tau }\pi $, (iv) $\tau \rightarrow \nu _{\tau }\rho 
$, (v) $\tau \rightarrow \nu _{\tau }a1$, and (vi) $\tau \rightarrow \nu
_{\tau }X$. These decays have branching ratios of 0.18, 0.18, 0.12, 0.26,
0.13, 0.13 \cite{pdg}, respectively. All the decays give a $\nu _{\tau }$.
The first two decays couple the tau propagation with $\nu _{e}$ and $\nu
_{\mu }$ propagation. The last decay includes the rest of the hadronic
decays not specified in (iii) through (vi). For a review, see \cite
{tkgaisser,plipari}.

\section{Results and Discussion for Fluxes}

Below we show results for total fluxes, including all the neutrino species
and their respective antineutrinos, unless defined otherwise. We plot upward
fluxes instead of up-to-down flux ratios to find the region in the $%
(E,\theta )$-space to compare the SM and LSG models in terms of absolute
flux differences. Larger fluxes mean more events, and larger difference in
the number of neutrinos in SM and LSG means better chances to differentiate
between the models. The ratio plots are not always helpful for that purpose
because they do not show us the actual number of neutrinos and the flux
difference of the SM and LSG. We present plots for $\Delta F_{1}(E,\theta )$%
, $\Delta F_{2}(E,\theta )$ in the $(E,\theta )$-space, where, 
\begin{equation}
\Delta F_{1}(E,\theta )=F_{SM}(E,\theta )-F_{LSG1}(E,\theta ),  \label{df1}
\end{equation}
and,

\begin{equation}
\Delta F_{2}(E,\theta )=F_{SM}(E,\theta )-F_{LSG2}(E,\theta ).  \label{df2}
\end{equation}
Here $F_{LSG1}(E,\theta )$, $F_{LSG2}(E,\theta )$ are the total upward
fluxes in the low scale gravity models with mass scale 1TeV and 2TeV
respectively, and number of extra dimensions $n=6$. $F_{SM}(E,\theta )$ is
the upward flux in standard model. $\Delta F_{1}(E),$ $\Delta F_{2}(E)$, $%
\Delta F_{1}(\theta )$, and $\Delta F_{2}(\theta )$ are defined in the same
manner as $\Delta F_{1}(E,\theta )$ and $\Delta F_{2}(E,\theta )$ are
defined above.

\begin{figure}
\includegraphics[width=2.3in,angle=-90]{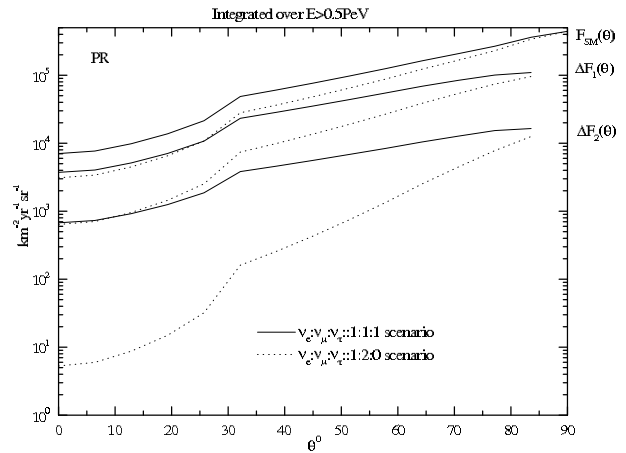} 
\caption{Upward neutrino flux ($km^{-2}.yr^{-1}.sr^{-1})$integrated over
energy $E>0.5PeV$ vs nadir angle $\theta $ (degrees) for R. J. Protheroe
input flux model: SM upward neutrino flux $F_{SM}(\theta )$ , and flux
differences $\Delta F_{1}(\theta ),$ and $\Delta F_{2}(\theta )$, as defined
in Eqs. \ref{df1} and \ref{df2}, are plotted for two scenarios: $\nu
_{e}:\nu _{\mu }:\nu _{\tau }::1:1:1$ (solid lines), and $\nu _{e}:\nu _{\mu
}:\nu _{\tau }::1:2:0$ (dotted lines).}
\label{fig:f4}
\end{figure}
\begin{figure}
\includegraphics[width=2.6in,angle=-90]{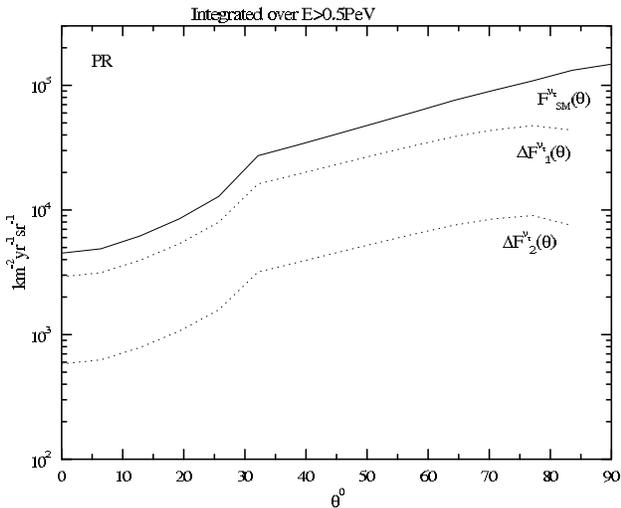}
\caption{Upward $\nu _{\tau } $ flux ($km^{-2}.yr^{-1}.sr^{-1})$ integrated
over energy $E>0.5PeV$ vs nadir angle $\theta $ (degrees) for R. J.
Protheroe input flux model: SM upward $\nu _{\tau }$ flux $F_{SM}^{\nu
_{\tau }}(\theta )$ (solid line), and $\nu _{\tau }$ flux differences $%
\Delta F_{1}^{\nu _{\tau }}(\theta ),$ and $\Delta F_{2}^{\nu _{\tau
}}(\theta )$ (dotted lines), as defined in Eqs. \ref{df1} and \ref{df2}, for
the scenarios: $\nu _{e}:\nu _{\mu }:\nu _{\tau }::1:1:1.$}
\label{fig:f5}
\end{figure}
\begin{figure}
\includegraphics[width=2.6in,angle=-90]{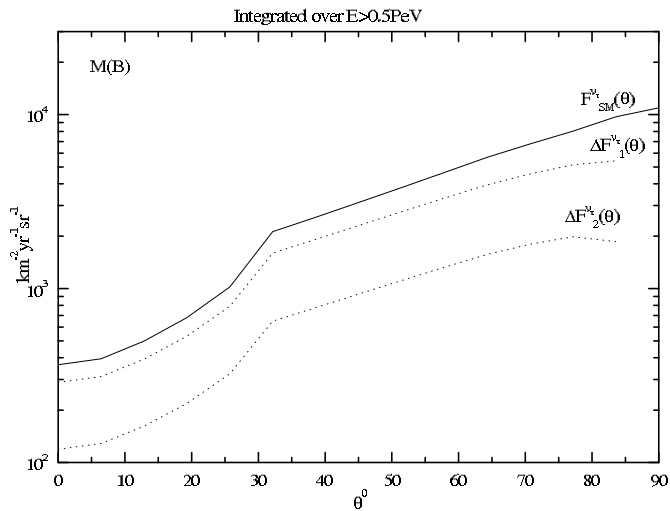} 
\caption{Same as Fig. 5, but with K. Mannheim (B) flux model.}
\label{fig:f6}
\end{figure}
\begin{figure}
\includegraphics[width=2.6in,angle=-90]{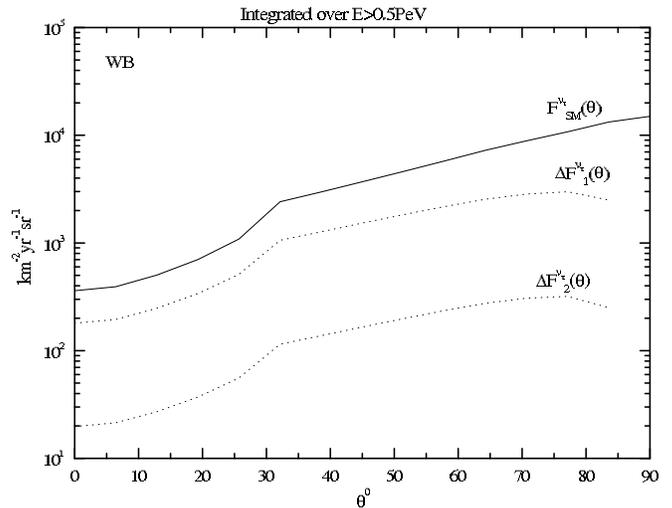}
\caption{Same as Fig. 5, but with Waxman Bahcall (WB) flux model.}
\label{fig:f7}
\end{figure}
\begin{figure}
\includegraphics[width=2.6in,angle=-90]{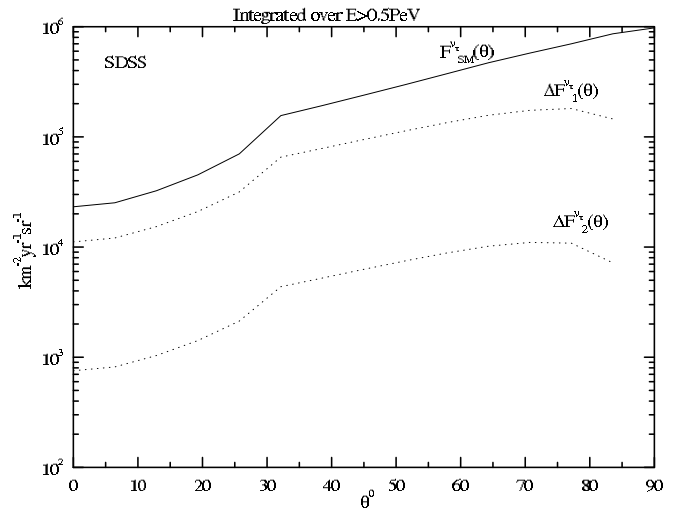}
\caption{Same as Fig. 5, but with \ SDSS flux model.}
\label{fig:f8}
\end{figure}
Below we show our analysis for the neutrino flux models due to R. J.
Protheroe \cite{rjproth}, K. Mannheim (B) \cite{manb}, Waxman Bahcall \cite
{wb}, SDSS \cite{sdss}, and 1/E generic model. Though we did not show it
here, we also looked at atmospheric and galactic neutrinos \cite{atmonutau}
around 0.5PeV . However, the up fluxes in this case are ignorable as
compared to the above extragalactic flux models: the galactic up flux is
more than an order of magnitude smaller than the up flux from any of the
flux models considered here; the atmospheric up flux is more than two orders
of magnitude lower than even the galactic up flux. The reason is quite
simple: one should expect larger feed down in the upward model fluxes
considered here; these extragalactic models have much larger fluxes above
1PeV than the atmospheric or galactic fluxes. As a result, the model fluxes
at higher energies give larger feed down at 0.5PeV while the atmospheric and
galactic up fluxes are so small there that they give essentially no feed
down. This is especially true in SM because of the strong feed down effect
per incident $\nu _{\tau }$ due to tau decays, which may make the
extragalactic model up fluxes, in SM, large enough to be detectable even at 0%
$^{0}$ nadir angles around 0.5PeV .

We will first show some plots (Figs. 4-14) to explore $(E,\theta )$-space of
the neutrino fluxes, and finally we will give two tables of numbers for
different neutrino flavor flux ratios. Three types of plots are shown below:

\textbf{(i) Fluxes integrated over energy vs nadir angle:} Figs. 4 and 5
refer to our example of R. J. Protheroe \cite{rjproth} model. Fig. 4 shows
the angular distribution of the total SM flux $F_{SM}(\theta )$ integrated
over energy $E>0.5PeV,$ along with the flux differences $\Delta F_{1}(E)$
and $\Delta F_{2}(E)$ as defined in the Eqs.\ref{df1} and \ref{df2}. The
qualitative features of the Figs.4 and 5 are the same in every flux model,
so we do not show figures for the other flux models corresponding to Fig.4.
However, we show the corresponding figures for Fig. 5 for the four models
(R. J. Protheroe \cite{rjproth}, K. Mannheim (B) \cite{manb}, Waxman Bahcall 
\cite{wb}, SDSS \cite{sdss}) because our emphasis here is on $\tau $%
-effects, and we wish to show their qualitative features are independent of
the flux model. As mentioned earlier, the atmospheric and galactic \cite
{atmonutau} up fluxes are not significant as compared to the four
extragalactic source models at these energies, so we do not consider them
here$.$

We plot the upward fluxes in two scenarios for the initial flux: $\nu _{e}$, 
$\nu _{\mu }$, $\nu _{\tau }::1$, $1$, $1$ (solid lines) and $\nu _{e}$, $%
\nu _{\mu }$, $\nu _{\tau }::1$, $2$, $0$ (dotted lines) . The two scenarios
correspond to $\nu _{\mu }\rightarrow \nu _{\tau }$ oscillations and no
oscillations in space, respectively. The following observations are common
to all models and are relevant to Fig. 4: (i) the difference is largest in
number around 75$^{0}$-85$^{0}$; however, as we will see in 3D plots in $%
(E,\theta )$-space, the maximum of the difference shifts to lower angles at
lower energies, (ii) around 30$^{0}$ nadir angle, one can clearly see the
effect of the core--- stronger suppression, (iii) the difference in flux
between SM and LSG models is larger at any angle in the $\nu _{e}$, $\nu
_{\mu }$, $\nu _{\tau }::1$, $1$, $1$ scenario than in the no tau scenario.
This is expected because of the stronger tau regeneration effect in the SM
as compared to LSG; in LSG, the black hole cross section, being the largest
of all as shown in Fig. 1, suppresses regeneration due to any process. This
observation leads us to concentrate on tau fluxes only, as discussed below.

In Figs. 5-8 we show the same plots for $\nu _{\tau }$ as was shown for the
total fluxes. We plot these figures for all the four models: R. J. Protheroe 
\cite{rjproth}, K. Mannheim (B) \cite{manb}, Waxman Bahcall \cite{wb}, and
SDSS \cite{sdss}. In these figures, the solid line is the $\nu _{\tau }$
flux in SM and the two dotted lines are the difference fluxes (see Eqs.\ref
{df1}, \ref{df2} ) for $\nu _{\tau }$. We clearly see that in LSG1 model,
regardless of the flux model, difference in $\nu _{\tau }$ fluxes is more
than 50 \% of the total difference due to all neutrino species. For example,
the $\nu _{\tau }$ difference $\Delta F_{1}(E)$ (Fig. 5) for Protheroe model
at 0$^{0}$ nadir angle is around $3000$, while the total difference is
around 4000 (Fig. 4), $km^{-2}yr^{-1}sr^{-1}.$ This behavior is independent
of the flux models--- about 3/4 of the total difference is due to $\nu
_{\tau }$ only. This may be useful as ICECUBE is expected to differentiate
between neutrino flavors around 500 TeV \cite{icecube}.

\textbf{(ii) Flux integrated over nadir angle vs energy: }The fluxes
integrated over nadir angle do not show as much detail as the ones
integrated over energy. This is because the integrated flux gets dominant
contribution from nadir angles around 90$^{0}$ where the chord length of the
earth is not long enough to make the difference between SM and LSG
prominent. For this reason, it suffices to show plots for only one flux
model , chosen to be the R. J. Protheroe model \cite{rjproth}; also the
qualitative features of these plots, like the earlier plots, are model
independent so it is not important to show the figures for all the models.
We see in Fig. 9: (a) Even at energies $E=0.5PeV$, the LSG1 and SM flux
models are distinguishable; (b) in the $\nu _{e}$, $\nu _{\mu }$, $\nu
_{\tau }::1$, $1$, $1$ scenario, $\Delta F_{1}(E)$ and $\Delta F_{2}(E)$ are
larger at lower energies, however, in contrast in the $\nu _{e}$, $\nu _{\mu
}$, $\nu _{\tau }::1$, $2$, $0$ scenario, they decrease with decreasing
energy. This contrasting behavior of the two scenarios begins around 1PeV
for LSG1 and around 5PeV for LSG2. This is expected because the stronger
feed down effect due to taus causes $F_{SM}(E)$ in the $\nu _{e}$, $\nu
_{\mu }$, $\nu _{\tau }::1$, $1$, $1$ scenario to increase faster with
decreasing energy than in the other scenario; however, $F_{LSG1}(E)$ and $%
F_{LSG2}(E)$ are not as different in the two scenarios because they are not
as sensitive to tau regeneration. Hence, the over all result is decreasing
flux difference for the $\nu _{e}$, $\nu _{\mu }$, $\nu _{\tau }::1$, $2$, $0
$ scenario and increasing flux difference for the other scenario. This
happens at a lower energy in the LSG1 model due to the lower energy scale
for LSG1 (Figs.1, 2); (c) another important observation is that $\Delta
F_{1}(E)$ is almost equal to $F_{SM}(E)$ after 3PeV, which means if we want
to differentiate the two models on the basis of the nadir angle integrated
event rates, the best region in energy may be around 3PeV, if the detector
has large enough efficiency to detect this flux. However, obtaining larger
fluxes for better statistics requires looking at lower energies. Though the
model fluxes around energies as low as 100 TeV are larger, the percent
difference between SM and LSG fluxes becomes smaller and smaller at energies
below 0.5PeV where SM cross sections are dominant; the total upward flux
rises much faster than the flux difference, making it hard to differentiate
between the two models. The atmospheric background is also larger at these
energies, hence we did not find it interesting to show the fluxes below
0.5PeV.

\begin{figure}
\includegraphics[width=2.6in,angle=-90]{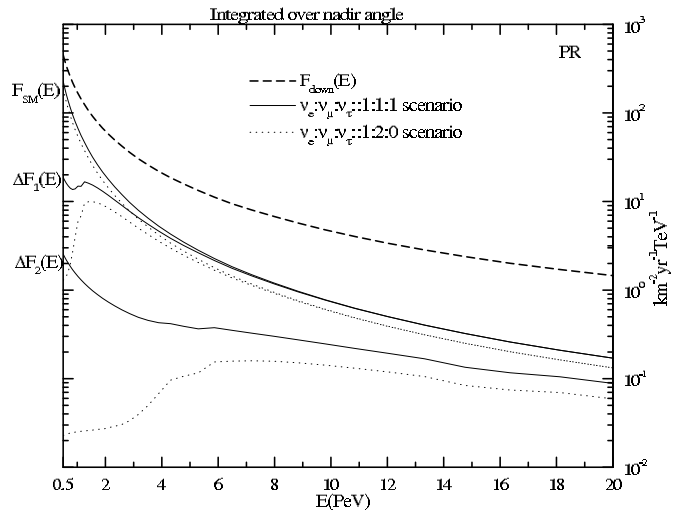}
\caption{Upward neutrino flux ($km^{-2}.yr^{-1}.TeV^{-1})$ integrated over
nadir angle vs energy $E$ ($PeV$) for R. J. Protheroe input flux model: SM
upward neutrino flux $F_{SM}(E)$ , and flux differences $\Delta F_{1}(E),$
and $\Delta F_{2}(E)$, as defined in Eqs. \ref{df1} and \ref{df2}, are
plotted for two scenarios: $\nu _{e}:\nu _{\mu }:\nu _{\tau }::1:1:1$ (solid
lines)$,$ and $\nu _{e}:\nu _{\mu }:\nu _{\tau }::1:2:0$ (dotted lines).
Also shown is the downward R. J. Protheroe model flux, integrated over nadir
angle (dashed line).}
\label{fig:f9}
\end{figure}
\begin{figure}
\includegraphics[width=2.6in,angle=-90]{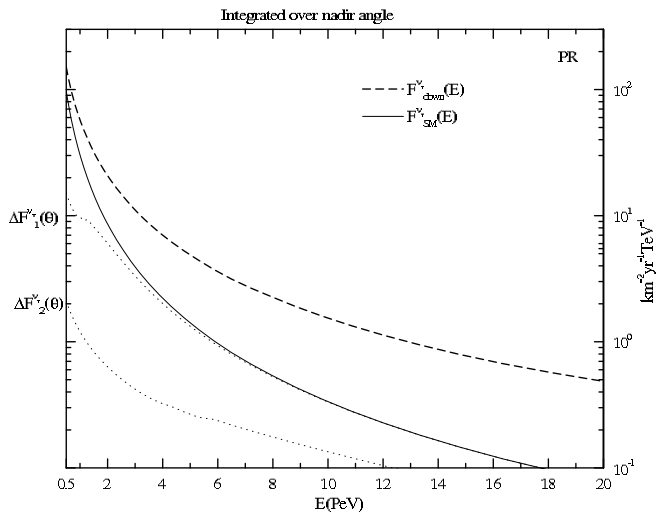}
\caption{Upward $\nu _{\tau }$ flux ($km^{-2}.yr^{-1}.TeV^{-1})$ integrated
over nadir angle vs energy $E$ ($PeV$) for R. J. Protheroe input flux model:
SM upward $\nu _{\tau }$ flux $F_{SM}^{\nu _{\tau }}(E)$ (solid line), and $%
\nu _{\tau }$ flux differences $\Delta F_{1}^{\nu _{\tau }}(E),$ and $\Delta
F_{2}^{\nu _{\tau }}(E)$ (dotted lines), as defined in Eqs. \ref{df1} and 
\ref{df2}, for the scenarios: $\nu _{e}:\nu _{\mu }:\nu _{\tau }::1:1:1.$
Also shown is the downward $\nu _{\tau }$\ R. J. Protheroe model flux,
integrated over nadir angle (dashed line).}
\label{fig:f10}
\end{figure}
In Fig. 10 we plot the nadir angle integrated flux of $\nu _{\tau }$ only.
If we compare Fig. 9 and 10, we come up with the similar answer as we did
for the flux integrated over energy ($\Delta F_{1}^{\nu _{\tau }}(E))$: the
upward $\nu _{\tau }$ flux difference, $\Delta F_{1}^{\nu _{\tau }}(\theta ),
$ around 0.5PeV is almost 3/4 of the total upward flux difference $\Delta
F_{1}(\theta )$. This again gives one hope that the signals of low scale
gravity may appear even around 0.5PeV.

\textbf{(iii) Plots of flux as a function of both energy and nadir angle: }%
Figs.11-14 give the complete detail of the fluxes in the $(E,\theta )$-space
for our flux example of R. J. Protheroe \cite{rjproth}. In Figs. 11 and 12
we plot the total upward flux in SM , $F_{SM}(E,\theta ),$ and the total
upward flux difference $\Delta F_{1}(E,\theta ),$ respectively; Figs. 13 and
14 have similar plots for the $\nu _{\tau }.$ We can see in these plots: (a)
The difference is the largest around 80$^{0}$, however, it is still
increasing even at 0.5PeV. Again, this may be surprising at first glance
because below 1PeV there is no significant contribution to the cross
sections from LSG. However, the reason is simply that the cross section at a
given energy will affect the neutrino flux at equal \textit{and lower}
energies due to feed down. Keeping this in mind, we can argue that the flux
around 1PeV or below gets more feed down from higher energies in SM because
LSG black hole cross section, being the largest of all the cross sections,
suppresses the feed down effect due to any process. (b) At higher energies
the flux difference peaks at higher angles e.g. around 80$^{\circ };$ the
peak shifts to lower angles at lower energies. One may argue that the peak
should always occur at the lowest nadir angle because the neutrinos will
have more interactions as they pass through earth with larger chord lengths,
and hence the SM and LSG models' interaction will cause the flux differences
to become larger and larger at lower angles and higher energies. However,
this does not happen because the input fluxes at higher energies are so
small that at lower nadir angles all the flux is either absorbed or fed down
to lower energies; that is why we see the flux difference peak shifts
towards 0$^{\circ }$ nadir angle at lower energies: feed down effect makes
the difference, between SM and LSG, at higher energies appear at lower
energies. (c) The $\nu _{\tau }$ plots in Figs. 13, 14 show us that the
major contributor of the difference between SM and LSG is the $\nu _{\tau }$
at lower energies. It contributes almost 3/4 of the total difference around
0.5PeV and around 1/3 at 10PeV. This feature is best seen in these full $%
(E,\theta )$-space plots. Around energies 0.5PeV, $\nu _{\tau }$ plays
important role in probing new physics. At energies around 10PeV and higher, $%
\nu _{\tau }$ behaves more like $\nu _{e}$ and $\nu _{\mu };$ this is
because around these energies the feed down due to taus, from even higher
energies, is not a big effect both in SM and LSG. This is a result of larger
tau decay lengths, smaller interaction lengths (see Figs. 1, 2), and smaller
fluxes at higher energies.

\begin{figure}
\includegraphics[width=2.6in,angle=-90]{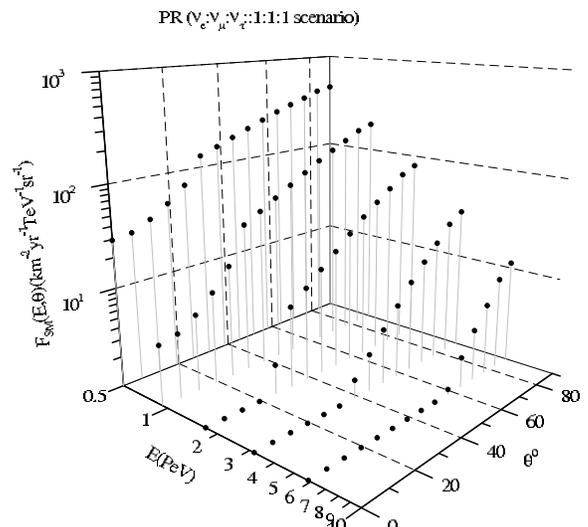}
\caption{SM upward neutrino flux $F_{SM}(E,\theta )$ ($%
km^{-2}.yr^{-1}.TeV^{-1}.sr^{-1})$ \ vs energy $E$ ($PeV$) \ and nadir angle 
$\theta (\deg ),$ for R. J. Protheroe input flux model in the scenario: $\nu
_{e}:\nu _{\mu }:\nu _{\tau }::1:1:1.$}
\label{fig:f11}
\end{figure}
\begin{figure}
\includegraphics[width=2.6in,angle=-90]{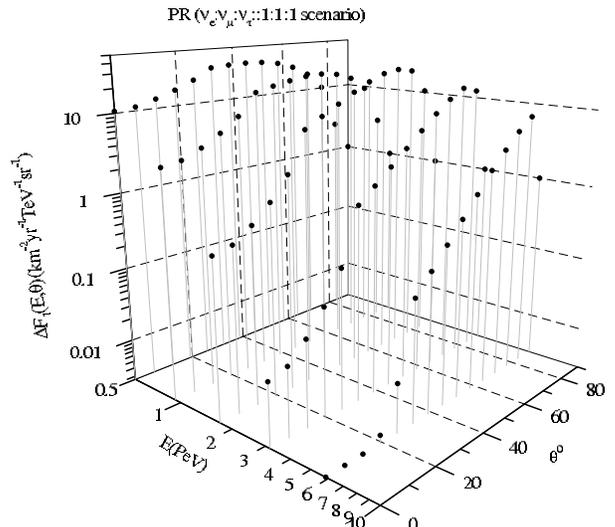}
\caption{Upward neutrino flux difference $\Delta F_{1}(E,\theta )$ ($%
km^{-2}.yr^{-1}.TeV^{-1}.sr^{-1})$ \ vs energy $E$ ($PeV$) \ and nadir angle 
$\theta (\deg ),$ for R. J. Protheroe input flux model in the scenario: $\nu
_{e}:\nu _{\mu }:\nu _{\tau }::1:1:1.$}
\label{fig:f12}
\end{figure}
Summarizing, we see that the above analysis discloses the flux structure $%
in (E,\theta )$-space: (i) Around 0.5PeV, the flux difference peaks in the
$40-60^{0}$ region. 
The larger angles tend to wash out the difference between LSG and
SM. (ii) Around 5PeV, the difference peaks in the
$75-80^{0}$ region. (iii) At higher energies, one will have to look at even
 larger nadir angles to get any detectable up flux. 
\begin{figure}
\includegraphics[width=2.6in,angle=-90]{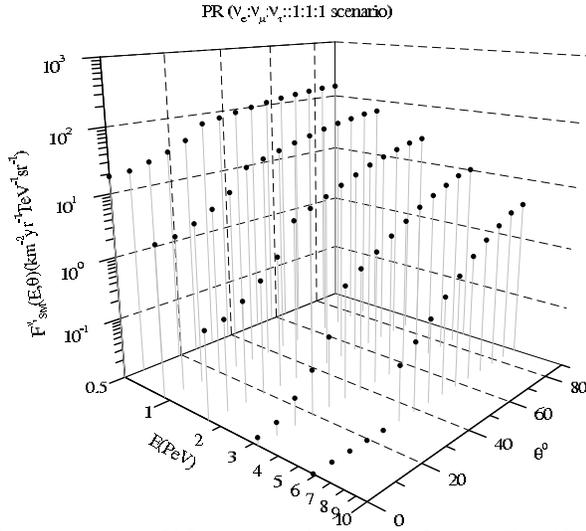}
\caption{SM upward $\nu _{\tau }$ flux $F_{SM}(E,\theta )$ ($%
km^{-2}.yr^{-1}.TeV^{-1}.sr^{-1})$ \ vs energy $E$ ($PeV$) \ and nadir angle 
$\theta (\deg ),$ for R. J. Protheroe input flux model in the scenario: $\nu
_{e}:\nu _{\mu }:\nu _{\tau }::1:1:1.$}
\label{fig:f13}
\end{figure}
\begin{figure}
\includegraphics[width=2.6in,angle=-90]{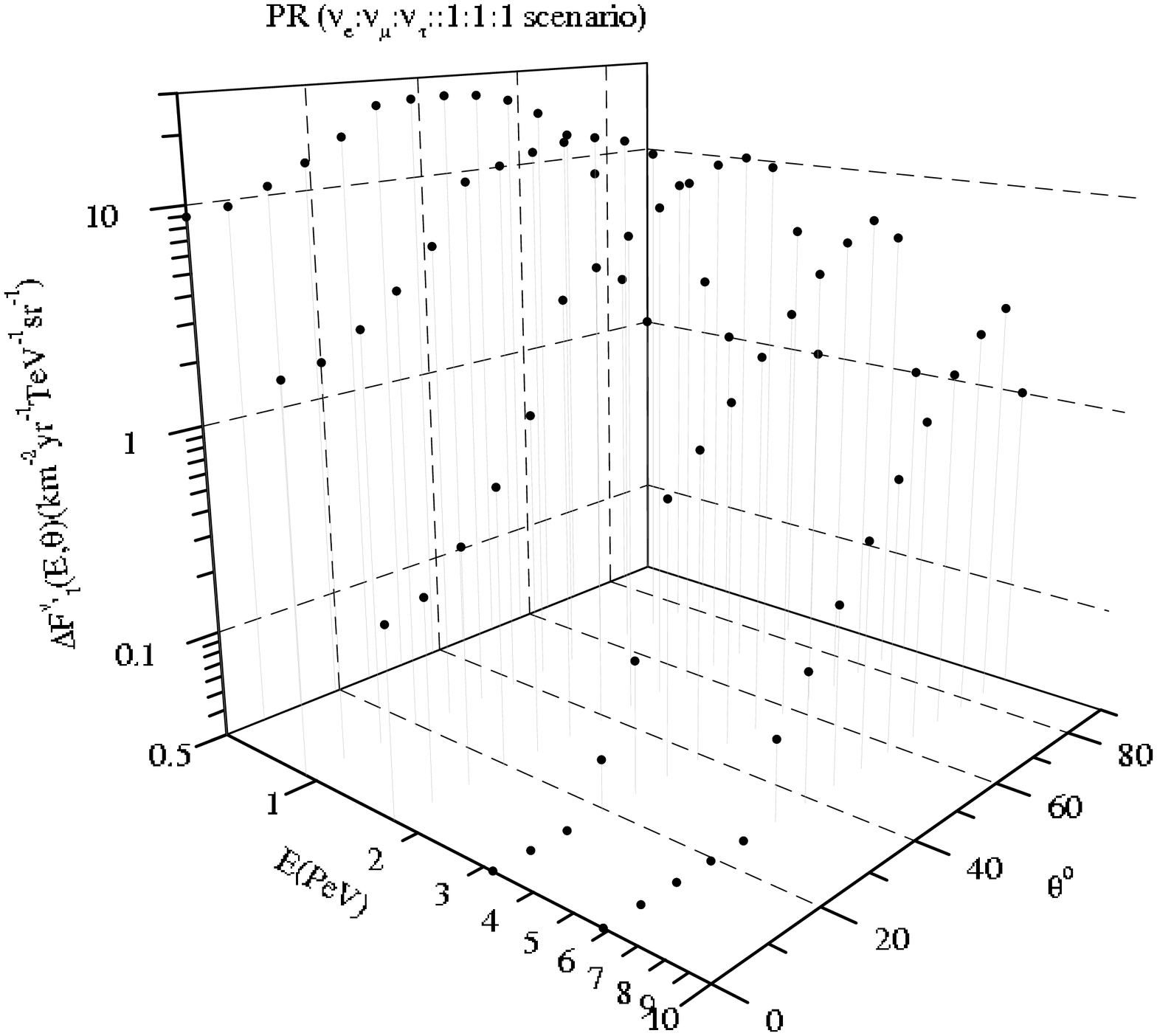}
\caption{Upward $\nu _{\tau }$\ flux difference $\Delta F_{1}^{\nu _{\tau
}}(E,\theta )$ ($km^{-2}.yr^{-1}.TeV^{-1}.sr^{-1})$ \ vs energy $E$ ($PeV$)
\ and nadir angle $\theta (\deg ),$ for R. J. Protheroe input flux model in
the scenario: $\nu _{e}:\nu _{\mu }:\nu _{\tau }::1:1:1.$}
\label{fig:f14}
\end{figure}
In Tables~\ref{tab:table1} and \ref{tab:table2}, we give different
flux ratios $R1$, $R2$, and $R3$ defines as, 
\begin{eqnarray}
&&R1=\frac{F^{\text{total}}\text{(up)}}{F^{\text{total}}\text{(down)}},\text{
}R2=\frac{F_{\nu _{\tau }}\text{(up)}}{F_{\nu _{\tau }}\text{(down)}},\text{ 
}  \nonumber \\
&&R3=\frac{F_{\nu _{\tau }}\text{(up)}}{F_{\nu _{e}}\text{(up)+}F_{\nu _{\mu
}}\text{(up)}},  \label{eqratio}
\end{eqnarray}
in the $\nu _{e}$, $\nu _{\mu }$, $\nu _{\tau }::1$, $1$, $1$ scenario, for
SM, LSG with M=2TeV (G2), and LSG with M=1TeV (G1). These ratios reveal some
interesting and useful features: (i) If we compare $R2$ and $R3$ with $R1$
in the Tables~\ref{tab:table1} and \ref{tab:table2}, we find that $\nu
_{\tau }$ flux indeed behaves differently in SM and LSG. (ii) In LSG, at
higher energies, $\nu _{e}$, $\nu _{\mu }$, and $\nu _{\tau }$ tend to
become identical as expected from larger decay length of taus at higher
energies and larger black hole cross sections hence smaller feed down of $%
\nu _{\tau }$ from tau decays; this effect is indeed seen for LSG (1TeV)
(see G1 in Table~\ref{tab:table2}); $R1$ and $R2$ tend to become equal and $%
R3$ tends to become 0.5 as expected (see G1 in Table~\ref{tab:table2}). The
same will be true for LSG (2TeV) at even higher energies.(iii) No matter
which neutrino flux model is correct, even at energies as low as 0.5PeV, we
see a clear difference between SM and LSG (1TeV) model (Table~\ref
{tab:table1}); at higher energies, $E>10PeV,$ there should be a difference
between LSG(2TeV) and SM based on their their comparison at 5PeV (Table~\ref
{tab:table2}). Our data at 10PeV which is not shown here, implies that,
based on the flux ratios, the \% difference between LSG(2TeV) and SM at
10PeV is bigger than the one between LSG (1TeV) and SM at 5PeV.(iv)
Isolating $\nu _{\tau }$ should help differentiate between SM and LSG, and
may help to differentiate between different neutrino flux models given SM
dynamics. For example, in Tables~\ref{tab:table1} and \ref{tab:table2},
though $R1$ is similar for different flux models in the SM case, $R3$ shows
some significant variation. 

\begin{table}
\caption{Flux ratios of the fluxes integrated over energy $E>0.5PeV$ at a
fixed nadir angle $\theta =45^{0}$, for SM, LSG with M = 2TeV ($G2$), and
LSG with M=1TeV ($G1$)$;$ $R1,$ $R2,$ and $R3$ are defined in Eqs. \ref
{eqratio}.}
\label{tab:table1}
\begin{tabular}{|c|c|c|c|}
\hline
MODEL & $R1$ & $R2$ & $R3$ \\ \hline
$
\begin{array}{l}
\\ 
WB \\ 
M(B) \\ 
PR \\ 
SDSS \\ 
1/E
\end{array}
$ & $
\begin{array}{lll}
SM & G2 & G1 \\ 
0.18 & 0.17 & 0.12 \\ 
0.15 & 0.12 & 0.06 \\ 
0.17 & 0.16 & 0.09 \\ 
0.18 & 0.18 & 0.12 \\ 
0.10 & 0.05 & 0.02
\end{array}
$ & $
\begin{array}{lll}
SM & G2 & G1 \\ 
0.25 & 0.23 & 0.14 \\ 
0.29 & 0.20 & 0.07 \\ 
0.27 & 0.24 & 0.12 \\ 
0.24 & 0.24 & 0.15 \\ 
0.22 & 0.09 & 0.02
\end{array}
$ & $
\begin{array}{lll}
SM & G2 & G1 \\ 
0.85 & 0.82 & 0.66 \\ 
1.61 & 1.26 & 0.73 \\ 
1.13 & 1.03 & 0.73 \\ 
0.83 & 0.82 & 0.66 \\ 
2.84 & 1.60 & 0.80
\end{array}
$ \\ \hline
\end{tabular}
\end{table}
\begin{table}
\caption{Flux ratios of the fluxes integrated over nadir angle $\theta $ at
a fixed energy $E=5PeV$ , for SM, LSG with M = 2TeV (columns $G2$), and LSG
with M = 1TeV (columns $G1$)$;$ $R1,$ $R2,$ and $R3$ are defined in Eqs. \ref
{eqratio}.)}
\label{tab:table2}
\begin{tabular}{|c|c|c|c|}
\hline
MODEL & $R1$ & $R2$ & $R3$ \\ \hline
$
\begin{array}{l}
\\ 
WB \\ 
M(B) \\ 
PR \\ 
SDSS \\ 
1/E
\end{array}
$ & $
\begin{array}{lll}
SM & G2 & G1 \\ 
0.20 & 0.18 & 0.02 \\ 
0.28 & 0.22 & 0.02 \\ 
0.21 & 0.19 & 0.02 \\ 
0.19 & 0.18 & 0.02 \\ 
0.43 & 0.24 & 0.02
\end{array}
$ & $
\begin{array}{lll}
SM & G2 & G1 \\ 
0.25 & 0.22 & 0.02 \\ 
0.45 & 0.30 & 0.02 \\ 
0.29 & 0.23 & 0.02 \\ 
0.23 & 0.20 & 0.02 \\ 
0.82 & 0.35 & 0.02
\end{array}
$ & $
\begin{array}{lll}
SM & G2 & G1 \\ 
0.71 & 0.65 & 0.53 \\ 
1.12 & 0.84 & 0.56 \\ 
0.80 & 0.69 & 0.54 \\ 
0.66 & 0.62 & 0.53 \\ 
1.72 & 0.93 & 0.57
\end{array}
$ \\ \hline
\end{tabular}
\end{table}

However, results of propagated WB and SDSS models are close to each other; 
they are similar
because, though WB down flux at lower energies is much smaller than SDSS, WB
flux falls like $E^{-2}$ while SDSS goes like $E^{-3}$ at higher energies
giving a stronger feed down effect for the former, hence, their up-down flux
ratios tend to be the same; the big difference between these models is that
WB has much weaker flux at energies 0.1-10PeV (Figs. 6 and 7 ). (v) The
difference among flux models are largely washed out by the LSG dynamics at
higher energies. For example, in Table~\ref{tab:table2}, $R1$ and $R2$ are
the same for all the flux models given here when the LSG scale is 1TeV. As
mentioned above, to probe higher LSG scales, one must go to higher energy
data which shows some sensitivity upto $M=5TeV$ \cite{prop2}.

\section{Event Rates}

Next we outline the formalism to calculate event rates for showers, muons, and
taus. Our formalism adds some refinement to the event rate estimates 
\cite{dbbeac,renoeve}. In presenting event rates, we take our theoretical 
"ICECUBE-like" detector to be 1 $km^3$ of strings of optical modules deployed with 
125m horizontal spacing. Because rates of events depend on flux and cross section, 
the extra depletion of upward flux in the LSG models is somewhat compensated by the 
increased  interaction probability of each neutrino that penetrates the detector 
effective volume. 
  
\textbf{(i) Shower rates: }Neutrino-nucleon interactions at PeV energies
and above initiate electromagnetic and hadronic showers which may produce a detectable
radio or optical Cherenkov signal in detectors like RICE, AMANDA, ICECUBE. For shower
rates in SM, we include both CC and NC interactions of $\nu _{e}$ and $\nu
_{\tau }$ but only NC interactions for $\nu _{\mu }$ (CC interaction in this
case gives muons which can be detected directly in detectors like AMANDA and
ICECUBE). For LSG, we include both eikonal and black hole cross sections.
For shower rates $R_{shower}$ in LSG, 
\begin{eqnarray}
R_{shower} &\cong&AL_{eff}^{s}\rho N_{A}\sum_{i}\int%
\limits_{E_{0}}^{\infty }dE_{\nu _{i}}^{p}F^{i}(E_{\nu _{i}}^{p}) 
( \sigma _{BH}(E_{\nu _{i}}^{p})\nonumber \\
&&+\int\limits_{E_{h0}/E_{\nu _{i}
}^{p}}^{1}dy\frac{d\sigma _{CC+NC+EK}^{i}(E_{\nu _{i}}^{p},y)}{dy})  
\end{eqnarray}
where $A$ is the detector area and $L_{eff}^{s}=$ $L_{D}+0.3(km)$, as
discussed below for shower rates, is the effective length of a detector of
instrumental length $L_{D}$. $E_{\nu _{i}}^{p}$ is the primary neutrino
energy, $E_{0}$ has to be greater than or equal to the minimum energy at 
which the flux is known, and $E_{h0}$ is the minimum energy of the
hadronic shower. The sum over `i' is to account for different neutrino flavors. With
the exception of $\nu _{e}$ CC interaction, for which we must set $E_{h0}=0$
for the reason given below, we have to set $E_{h0}=E_{0}\geq 1TeV$ as we do
not know the flux below $E_{0}$ and hence cannot account for all the showers
produced below $E_{0}$. One needs $E_{h0}>1 TeV$ for the showers to be
detectable. For the black hole we assume that it has equal probability of
decaying into any SM particles, and  that it will always give a shower of
energy around $E_{0}$ and higher.

Two important points: \textbf{(i)} 
\textit{We take the effective detector length }$L_{eff}^{s}$\textit{\ as the
instrumental detector length }$L_{D}$\textit{\ plus 0.3km. This is because,
in addition to the showers produced inside the detecor, a conservative 
estimate is that shower signal
produced 0.15km outside the detector, on any side, will easily reach the
detector using a shower range of 0.3km for optical modules}. 
This increases the shower rates by 30\%.\textbf{(ii)} \textit{For
showers from }$\nu _{e}$\textit{\ CC interaction, there is no need to set a
lower limit on y as the electron energy will add to the hadron energy to 
contribute to the shower signal (e.g. 
effectively one
can set  }$E_{h0}=0$\textit{)}. This increases the shower rates dramatically
---30-50\%. This is because in this case we can set minimum y=0 and the CC
cross section peaks around y=0 giving a large percentage of the total shower
events. 

\textbf{(ii) Muon rates: }$\nu _{\mu }$ CC interaction and $\nu _{\tau }$ CC
interaction, with the tau decay $\tau \rightarrow \nu _{\tau }\nu _{\mu }\mu 
$, both are the sources of muons\footnote{%
We do not include here the taus from $\nu _{\tau }$ CC interaction that will
be mistaken by the detector as muons.}. For the former case, muon rate $%
R_{\mu }^{CC}$ is given by
\begin{eqnarray}
R_{\mu }^{CC} &\cong &A\rho N_{A}\int\limits_{E_{0}}^{\infty }dE_{\nu _{\mu
}}^{p}F^{\nu _{\mu }}(E_{\nu _{\mu }}^{p})\int\limits_{0}^{1-\frac{E_{\mu 0}%
}{E_{\nu_{\mu }}^{p}}}dy\frac{d\sigma _{CC}^{\nu _{\mu }}(E_{\nu _{\mu }}^{p},y)}{%
dy}  \label{rmcc} \nonumber \\
&&\text{ }L_{eff}^{\mu }(E_{\nu _{\mu }}^{p},y)\theta (\frac{R(E_{\mu
},E_{\mu 0})}{\rho }-(x_{\min }^{\mu track}+l^{show})),\nonumber \\  
\end{eqnarray}
where $E_{\mu 0}=E_{0}$ for the same reason as given above for the shower
rate. The $x_{\min }^{\mu track}\simeq 0.25km$ is the minimum muon track
length required to detect a muon and $l^{show}$ $\simeq $ $0.02km$ ($\ll $ $%
x_{\min }^{\mu track}$) is the typical shower size (for PeV energies) of the
shower produced at the $\nu _{\mu }$ event vertex. The $\theta $ function
guarantees the exclusion of muons whose range is too small for them to be
detected. $L_{eff}^{\mu }(E_{\nu _{\mu }}^{p},y)$ is the effective detector
length for muon rates: 
\begin{eqnarray}
L_{eff}^{\mu }(E_{\nu _{\mu }}^{p},y)&=&\frac{R(E_{\mu }(E_{\nu _{\mu
}}^{p},y),E_{\mu 0})}{\rho }+L_{D}-2x_{\min }^{\mu track}\nonumber \\ 
&&-l^{show} , \label{lmueff}
\end{eqnarray}
where 
\begin{equation}
\frac{R(E_{\mu },E_{\mu 0})}{\rho }=\frac{1}{\rho \beta }\ln \left( \frac{%
\alpha +\beta E_{\mu }}{\alpha +\beta E_{\mu 0}}\right)   
\end{equation}
is the average electromagnetic range, in a matter of density $\rho $, of a
muon of initial and final energies $E_{\mu }$ and $E_{\mu 0}$, respectively
Here $\alpha =2.0MeVcm^{2}/g$ accounts for the muon energy loss due to
ionization and $\beta =4.2\times 10^{-6}cm^{2}/g$ is due to pair production,
bremmstraulung, and photonuclear energy losses \cite{renomu}. 

We should emphasize that\textit{\ the effective detector length }$%
L_{eff}^{\mu }(E_{\nu _{\mu }}^{p},y)$\textit{\ given above is the appropriate
one for this case.} Our definition of the effective detector length works
for any value of the muon range while $\frac{R(E_{\mu },E_{\mu 0})}{\rho }$
works only for $\frac{R(E_{\mu },E_{\mu 0})}{\rho }\gg L_{D}$. 

Muon rate from the tau decay $\tau \rightarrow \nu _{\tau }\nu _{\mu }\mu $
is given by 
\begin{widetext}
\begin{eqnarray}
R_{\mu }^{\tau \rightarrow \mu } &\cong &A\rho
N_{A}\int\limits_{E_{0}}^{\infty }dE_{\nu _{\tau }}^{p}F^{\nu _{\tau
}}(E_{\nu }^{p})\int\limits_{0}^{1-\frac{E_{\tau 0}}{E_{\nu _{\tau }}^{p}}}dy%
\frac{d\sigma _{CC}^{\nu _{\tau }}(E_{\nu _{\tau }}^{p},y)}{dy} 
\int\limits_{0}^{\infty }dx\frac{e^{-x/\mathcal{L}_{dec}^{\tau }(E_{\tau
})}}{\mathcal{L}_{dec}^{\tau }(E_{\tau })}\int\limits_{E_{\mu 0}/E_{\tau
}}^{1}dz^{\prime }\frac{dP^{\tau \rightarrow \mu }(z^{\prime })}{dz^{\prime }%
}  \nonumber \\
&&L_{eff}^{\mu }(E_{\nu _{\tau }}^{p},z^{\prime })\theta \left( (\frac{%
R(z^{\prime }E_{\tau },E_{\mu 0})}{\rho })-(x_{\min }^{\mu
track}+l^{show})\right),
\end{eqnarray}
\end{widetext}
where $\mathcal{L}_{dec}^{\tau }$ is defined in Eq.\ref{ltdec}; $E_{\tau
0}=E_{0}$ for the same reason as given above for the shower and muon rate; $%
\frac{dP^{\tau \rightarrow \mu }(z^{\prime })}{dz^{\prime }}$ gives the
relevant decay distribution with $E_{\mu }=z^{\prime }E_{\tau }=z^{\prime
}(1-y)E_{\nu _{\tau }}^{p}$ \cite[2000]{smpreno}; $x_{\min }^{\mu track}$
and $l^{show}$ are the same as defined above. $L_{eff}^{\mu }$ is defined in
Eq. \ref{lmueff}. The integration over $x$ gives total probability that a tau
will decay with a tau decay length between 0 and $\infty $. The $\theta $ 
function requires the muon range to be greater than the
minimum distance required to detect a muon.

\textbf{(iii) Tau rates: }We discuss two types of events that are unique to
the presence of taus and $\nu _{\tau }$ \cite{dblp,dbap,dbah,dbbeac}:

(1) A tau produced in a $\nu _{\tau }$ CC interaction \textit{outside }the
detector decays (excluding the decay $\tau \rightarrow \nu _{\tau }\mu \nu
_{\mu }$)\textit{\ inside} the detector a track and a shower:
\begin{widetext}
\begin{eqnarray}
R_{\tau }^{\tau dec}(\text{1 shower}) &\cong &0.83A\rho
N_{A}\int\limits_{E_{0}}^{\infty }dE_{\nu _{\tau }}^{p}F^{\nu _{\tau
}}(E_{\nu _{\tau }}^{p})\int\limits_{0}^{1-\frac{E_{\tau 0}}{E_{\nu _{\tau
}}^{p}}}dy\frac{d\sigma _{CC}^{\nu _{\tau }}
(E_{\nu _{\tau }}^{p},y)}{dy}\nonumber \\
&&\left( \int\limits_{x_{\min }^{\tau
track}+0.15(km)}^{L_{D}+0.3(km)}dxL_{1eff}^{\tau }(x)\frac{e^{-x/\mathcal{L}%
_{dec}^{\tau }(E_{\tau })}}{\mathcal{L}_{dec}^{\tau }(E_{\tau })}%
+(L_{2eff}^{\tau }\int\limits_{L_{D}+0.3(km)}^{\infty }dx\frac{e^{-x/%
\mathcal{L}_{dec}^{\tau }(E_{\tau })}}{\mathcal{L}_{dec}^{\tau }(E_{\tau })}%
)\right) ,  
\end{eqnarray}
\end{widetext}
where
\begin{eqnarray}
L_{1eff}^{\tau }(x) &=&(x-x_{\min-0.15(km)}^{\tau track})\nonumber \\
L_{2eff}^{\tau } &=&(L_{D}+0.15(km)-x_{\min }^{\tau track}),  
\end{eqnarray}
where $E_{\tau 0}=E_{0}$ for the same reason as given above for the shower
and muon rate. The 0.83 factor is to exclude the decay $\tau \rightarrow \nu
_{\tau }\mu \nu _{\mu }$ which has a branching ratio of \symbol{126}0.17, and 
$x_{\min }^{\tau track}\simeq 0.25(km)$ is the minimum tau track length
required to detect a tau. We have assumed that all the showers produced in
these events will be detectable, so we do not need tau decay distribution
function. This is a reasonable assumption as we will choose $E_{0}=E_{\tau
0}=0.5PeV$ which means almost all of the showers produced from tau decay
will be above the detector threshold of $\sim 0.001PeV$. The lower
limit for x integration assures the tau decay length large enough for the
tau to be separately detected from the shower. The expression given for
these events in Ref.\cite{dbbeac} includes some $shower-track-shower$ 
events
too.  The expression above gives \textit{only} $track-shower$ events by using the $x$
dependent effective length where  $x$ is smaller than 
$L_{D}+0.3(km)$. Moreover, we have included the  0.3(km) in the $x$ integration limit and
0.15(km) in the expression for $L_{2eff}^{\tau }$. 
These numbers follow from the reasoning given above in discussion of shower
rates. 

(2) A tau produced in a $\nu _{\tau }$ CC interaction \textit{inside }the
detector decays (excluding the decay $\tau \rightarrow \nu _{\tau }\mu \nu
_{\mu }$)\textit{\ inside} the detector giving a shower, a track, and another shower. 
These are so called double bang
events \cite{dblp}:
\begin{widetext}
\begin{eqnarray}
R_{\tau }^{\tau dec}(\text{2 shower}) &\cong&0.83A\rho
N_{A}\int\limits_{E_{0}}^{\infty }dE_{\nu _{\tau }}^{p}F^{\nu _{\tau
}}(E_{\nu _{\tau }}^{p})
\int\limits_{0}^{1-\frac{E_{\tau 0}}{E_{\nu _{\tau
}}^{p}}}dy\frac{d\sigma _{CC}^{\nu _{\tau }}
(E_{\nu _{\tau }}^{p},y)}{dy} 
\int\limits_{x_{\min }^{\tau track}+l^{show}}^{L_{D}+0.3(km)}
dx\frac{e^{-x/\mathcal{L}_{dec}^{\tau }(E_{\tau })}}
{\mathcal{L}_{dec}^{\tau }(E_{\tau })}L_{3eff}^{\tau }(x),
\end{eqnarray}
\end{widetext}
where
\begin{equation}
L_{3eff}^{\tau }(x) =(L_{D}+0.3(km)-x) 
\end{equation}
and all the other symbols have been defined above.
\begin{table}
\caption{\label{tab:table3}Up and down events ($yr^{-1}$) for $E\geq 0.5PeV$ 
in the scenario 1:1:1; 
all upward events are integrated over nadir angle $\theta \leq 84^{0}$; down showers 
are integrated over angle  but muons and taus are not (see text for details).}
\begin{tabular}{|c|c|c|c|}
\hline
& showers ($\frac{up}{down}$) & muons ($\frac{up}{down}$) & taus ($\frac{up}{%
down}$) \\ \hline
$
\begin{array}{l}
\\ 
WB\smallskip \\ 
SD \smallskip \\ 
MB \smallskip \\ 
PR\smallskip 
\end{array}
$ & $
\begin{array}{lll}
SM & G2 & G1 \\ 
\frac{2.6}{11}\smallskip  & \frac{2.7}{24} & \frac{3.1}{201} \\ 
\frac{163}{622}\smallskip  & \frac{167}{748} & \frac{202}{4725} \\ 
\frac{3.0}{30}\smallskip  & \frac{3.1}{195} & \frac{2.2}{1898} \\ 
\frac{32}{182}\smallskip  & \frac{33}{534} & \frac{34}{5284}
\end{array}
$ & $
\begin{array}{lll}
SM & G2 & G1 \\ 
\frac{3.0}{3.3}\smallskip  & \frac{2.7}{3.3} & \frac{1.1}{3.3} \\ 
\frac{176}{142}\smallskip  & \frac{165}{142} & \frac{74}{142} \\ 
\frac{6.4}{18}\smallskip  & \frac{3.6}{18} & \frac{0.66}{18} \\ 
\frac{50}{73}\smallskip  & \frac{38}{73} & \frac{12}{73}
\end{array}
$ & $
\begin{array}{lll}
SM & G2 & G1 \\ 
\frac{0.11}{0.13}\smallskip  & \frac{.074}{0.13} & \frac{0.0086}{0.13} \\ 
\frac{4.7}{4.6}\smallskip  & \frac{4.0}{4.6} & \frac{0.54}{4.6} \\ 
\frac{0.46}{0.86}\smallskip  & \frac{0.20}{0.86} & \frac{0.01}{0.86} \\ 
\frac{2.4}{3.5} \smallskip & \frac{1.5}{3.5} & \frac{0.13}{3.5}
\end{array}
$ \\ \hline
\end{tabular}
\hspace{0pt}
\end{table}

In addition to the above events for tagging taus, we have looked at the
possibilty of detecting taus from the decay $\tau \rightarrow \nu _{\tau
}\mu \nu _{\mu }$ provided the tau decays \textit{inside} the detector. These
events have smaller rate than double bang or single shower events. Details of
this calculation will be given elsewhere.

\section{Results and Discussion for Event Rates}

The results for event rates are summarized in the Tables~\ref{tab:table3}-
\ref{tab:table6}. The event
rates at angles below 60$^{0}$ turn out to be very small for current
detectors, hence, we show events integrated to $\theta =84^{0}$ nadir angle.

In Table~\ref{tab:table3} we give different upward and down events 
per year, in the scenario 1:1:1, for a $1km^3$ detector in ice. Our down shower events 
have been integrated over angle, however, the down tau and down muon events 
correspond to near horizon events.
The showers contain events due to all neutrino flavors. These shower events
in 1:1:1 scenario are 30- 40\% larger than 1:2:0 scenario (not shown here).
This is because, in the latter, we are excluding $\nu _{\mu }$ CC interaction
and also the total flux is smaller due to the absence of tau regeneration
effect. Upward muons (muon up) contain muons from $\nu _{\mu }$ CC
interaction and the muonic tau decay. The muons in 1:1:1 scenario are about
40\% smaller than 1:2:0 scenario (not shown). They are not exactly 50\% of
the latter due to the contribution from tau decays. The upward taus (taus up)
contain all three types of  events described above (e.g. tau up =  
$(track-shower) +
(shower-track-shower) + 
\tau \rightarrow \nu _{\tau }\mu \nu _{\mu }$). The ratio
$(track-shower)/(shower-track-shower)$, 
not shown here, is very sensitive to the flux model
and can be anywhere between 0.7 to 1.7. The $\tau \rightarrow \nu _{\tau
}\mu \nu _{\mu }$ tau events are always the smallest ---less than 50\% of
the smaller of the single shower and double bang. While muons show differences
between LSG (1TeV)and SM, the up-to-down showers  and tau event ratios 
show a clearer differentiation between the two (Table~\ref{tab:table3}). The
number of events, though marginal for $\nu_{\tau}$'s in WB and MB, are sufficient to
make a clear distinction from SM for LSG (1 TeV) and distinction in some cases for 
LSG (2 TeV) in several years of data taking.  
The up tau events in the LSG (1 TeV) are especially severely suppressed 
as compared to the
SM. This again reflects the fact that tau decay is playing much weaker
role in LSG (1TeV) (see Section IV). However, as expected, taus play a similar role in SM and
LSG(2TeV), though some suppression of $\nu_{\tau}$ is evident for LSG (2TeV) 
in Tables~\ref{tab:table3} and~\ref{tab:table4}. In Table~\ref{tab:table4}, we see the LSG/SM ratio of the ratios, 
as defined in the table caption,
strongly differentiates SM from LSG(1TeV)in both of the flavor scenarios. 
\textit{In fact, these ratios have very weak dependence on the flavor scenario.} They 
are spectacularly large in every flux model. Even LSG (2TeV) is clearly distinguished in 
all but the SDSS model. In the latter, the number of events is large enough that one 
may hope to discriminate between LSG (2TeV) and SM.

We show similar tables with an energy threshold at 5PeV 
(Tables~\ref{tab:table5} and ~\ref{tab:table6}). Here we see the tau events did not 
change much in SM and LSG (2 TeV) model. However, in LSG (1TeV)up tau events have 
decreased by an order of magnitude, generally with enough down events to 
make them even more useful in differentiating 
between the two at 5PeV as compared to 0.5PeV. Although 
not shown here, by comparing 5PeV and 0.5PeV results for showers and muons 
one expects that LSG(2 TeV) model around 10PeV thresholds will differ 
from SM to the extent that LSG (1 TeV) does from SM around 0.5PeV. However, the event 
rates may be too small to do a statistical analysis.

Looking at the event rates for different flux models in Table~\ref{tab:table3}, we see the
PR and SDSS flux models provide large enough events to do a statistical analysis. For 
WB and MB models, only the down shower rates in LSG(1TeV) are large enough to differentiate it
from the SM by looking at the up-to-down shower ratios.

\begin{table}

\caption{\label{tab:table4}Ratios of the ratios; here 
$RR1$=$\frac{showers\text{ }down}{muons\text{ }up}$ and 
$RR2$=$\frac{taus\text{ }down}{taus\text{ }up}$.}
\begin{tabular}{|c|c|c|c|c|}
\hline

&$\frac{RR1_{G2}}{RR1_{SM}}$ & $\frac{RR1_{G1}}{RR1_{SM}}$ & 
$\frac{RR2_{G2}}{RR2_{SM}}$ & $%
\frac{RR2_{G1}}{RR2_{SM}}$ \\ \hline
$
\begin{array}{l}
\\
WB\smallskip \\ 
SD \smallskip \\ 
MB \smallskip \\ 
PR\smallskip 
\end{array}
$ & $
\begin{array}{ll}
1:2:0 & 1:1:1 \\ 
\frac{4.5}{1.7}\text{=2.7} \smallskip & \frac{8.8}{3.6}\text{=2.4} \\ 
\frac{2.1}{1.6}\text{= 1.3} \smallskip & \frac{4.5}{3.5}\text{=1.3} \\ 
\frac{30}{2.1}\text{= 14} \smallskip & \frac{54}{4.7}\text{=11} \\ 
\frac{7.4}{1.6}\text{=4.7} \smallskip & \frac{14}{3.6}\text{=3.9}
\end{array}
$ & $
\begin{array}{ll}
1:2:0 & 1:1:1 \\ 
\frac{93}{1.7}\text{=56} \smallskip & \frac{183}{3.6}\text{=50} \\ 
\frac{34}{1.6}\text{= 21} \smallskip & \frac{64}{3.5}\text{=18} \\ 
\frac{1571}{2.1}\text{= 735}\smallskip  & \frac{2875}{4.7}\text{=610} \\ 
\frac{249}{1.6}\text{=157}\smallskip  & \frac{440}{3.6}\text{=121}
\end{array}
$ & $
\begin{array}{l}
\\
1.5 \smallskip \\ 
 1.2 \smallskip \\ 
2.3 \smallskip \\ 
1.6\smallskip 
\end{array}
$ & $
\begin{array}{l}
 \\ 
13\smallskip  \\ 
 8.7 \smallskip \\ 
 46 \smallskip \\ 
18\smallskip 
\end{array}
$ \\ \hline
\end{tabular}
\end{table}

\begin{table}
\caption{\label{tab:table5}Same as Table~\ref{tab:table3} but with energy threshold 5PeV.}
\begin{tabular}{|c|c|c|c|}
\hline
& showers ($\frac{up}{down}$) & muons ($\frac{up}{down}$) & taus ($\frac{up}{%
down}$) \\ \hline
$
\begin{array}{l}
\\ 
WB\smallskip \\ 
SD \smallskip \\ 
MB \smallskip \\ 
PR\smallskip 
\end{array}
$ & $
\begin{array}{lll}
SM & G2 & G1 \\ 
\frac{0.26}{2.9}\smallskip  & \frac{0.33}{13} & \frac{0.15}{151} \\ 
\frac{11}{94}\smallskip  & \frac{15}{188} & \frac{8.7}{2714} \\ 
\frac{1.2}{21}\smallskip  & \frac{1.4}{157} & \frac{0.41}{1602} \\ 
\frac{6.5}{82} \smallskip & \frac{8}{371} & \frac{2.9}{4178}
\end{array}
$ & $
\begin{array}{lll}
SM & G2 & G1 \\ 
\frac{0.3}{0.84}\smallskip  & \frac{0.14}{0.84} & \frac{.002}{0.84} \\ 
\frac{8.4}{14}\smallskip  & \frac{5.6}{14} & \frac{0.1}{14} \\ 
\frac{2.0}{9.2}\smallskip  & \frac{0.66}{9.2} & \frac{.006}{9.2} \\ 
\frac{8.5}{26} \smallskip & \frac{3.7}{26} & \frac{.04}{26}
\end{array}
$ & $
\begin{array}{lll}
SM & G2 & G1 \\ 
\frac{0.07}{0.11}\smallskip  & \frac{0.04}{0.11} & \frac{.001}{0.11} \\ 
\frac{2.5}{3.1}\smallskip  & \frac{1.8}{3.1} & \frac{0.05}{3.1} \\ 
\frac{0.41}{0.83}\smallskip  & \frac{0.16}{0.83} & \frac{.003}{0.83} \\ 
\frac{1.9}{3.0} \smallskip & \frac{0.93}{3.0} & \frac{0.02}{3.0}
\end{array}
$ \\ \hline
\end{tabular}
\hspace{0pt}
\end{table}

\begin{table}
\caption{\label{tab:table6}Same as Table~\ref{tab:table4} but with energy threshold 5PeV.}
\begin{tabular}{|c|c|c|c|c|}
\hline
&$\frac{RR1_{G2}}{RR1_{SM}}$ & $\frac{RR1_{G1}}{RR1_{SM}}$ & $\frac{RR2_{G2}}{RR2_{SM}}$ & $%
\frac{RR2_{G1}}{RR2_{SM}}$ \\ \hline
$
\begin{array}{l}
\\ 
WB\\ 
SD  \\ 
MB  \\ 
PR
\end{array}
$ & $
\begin{array}{ll}
1:2:0 & 1:1:1 \\ 
 11 & 9.3 \\ 
3.2 & 3.0 \\ 
27 & 23 \\ 
12 & 10
\end{array}
$ & $
\begin{array}{ll}
1:2:0 & 1:1:1 \\ 
8400 & 7826 \\ 
2481 & 1419 \\ 
3.0e4 & 2.7e4
\\ 
1.2e4 & 1.0e4
\end{array}
$ & $
\begin{array}{l}
\\ 
1.7 \\ 
1.4 \\ 
2.6 \\ 
2.0
\end{array}
$ & $
\begin{array}{l}
 \\ 
 74 \\ 
 48 \\ 
 161 \\ 
100
\end{array}
$ \\ \hline
\end{tabular}
\end{table}

\section{Summary of Results and Conclusion}

We found complete numerical solutions to the system of coupled equations
that include the most important effects for transport through Earth of $\nu
_{e},\nu _{\mu },\nu _{\tau }$ and $\tau $ fluxes above 0.5 PeV. In Fig. 4,
we presented results of angular distributions of total neutrino flux in our
example of the diffuse flux model by R. J. Protheroe \cite{rjproth},
however, the qualitative features of this figure are common to the other
flux models (K. Mannheim (B) \cite{manb}, Waxman Bahcall \cite{wb}, and SDSS 
\cite{sdss}). Fluxes in this figure are integrated from 0.5 PeV upward,
showing the effects of including low scale gravity enhancement to the lepton
deep inelastic cross sections, \textit{with no $\nu _{\tau }$ and full $\nu
_{\tau }$ mixing into the incident flux.} This figure also show that the $%
\nu _{\tau }$ regeneration from $\tau $ decay enhances the ``through
earth'', or ``upward'' fluxes significantly more in the standard model than
in the models with low scale gravity enhancements included, as seen on the
curves where $\nu _{\tau }$ is mixed into the flux incident on Earth. The
standard model flux is obviously higher at nadir angles smaller than 80
degrees, while the differences between the fluxes with standard model
interactions only and those with low scale gravity included are much larger
at small nadir angles in the case that $\nu _{\tau }$ is mixed into the
incident flux than in the case when there it is not. Next in Figs. 5-8, we
showed the equivalent angular distributions for the $\nu _{\tau }$ flux
alone to emphasize the observation just summarized, that is, as compared to $%
\nu _{e}$ and $\nu _{\mu },$ $\nu _{\tau }$ can serve better to
differentiate between SM and LSG at energies below 10PeV.

As established by the angular distribution graphs, the qualitative features
are shared by all the models, so we gave only the Protheroe model results in
plotting the energy distribution of flux integrated over angles in the range
from 0.5 PeV to 20 PeV in Figs. 9 and 10. These plots show that in this
energy range, the low scale gravity interactions rapidly suppress the upward
flux compared to the standard model. They also indicate the fact that the
regeneration of $\nu _{\tau }$ flux is much less significant when low scale
gravity is turned on, as clearly indicated by the flux difference curves in
the energy range between 0.5 PeV and 2.0 PeV.

In the series of graphs from Figs.11 through 14, we displayed the
three-dimensional plots of the total and $\nu _{\tau }-only$ fluxes for the
standard model and for the low scale gravity, M = 1 TeV case. These indicate
in detail where the maximum flux differences are in angle and energy.

Next we looked at the flux ratios $R1,$ $R2,$ and $R3$ as defined in Eqs. 
\ref{eqratio}. The results are given in Tables~\ref{tab:table1} and \ref
{tab:table2}. Here again the distinction between $\nu _{\tau }$ and 
$\nu _{\mu }+\nu _{e}$ fluxes is evident. The
distinctions among flux models are largely washed out by the LSG dynamics at
higher energies. For example, in Table~\ref{tab:table2}, $R1$ and $R2$ are
the same for all the flux models given here.

In Sections V and VI we presented the defining equations for our shower, muon and 
tau rates and the results of our rate calculation. The story is summarized in Tables III-VI. 
Using a cutoff of 0.5PeV, we found that the events rates in showers and muon categories 
are large enough to make meaningful statements about the distinction between SM and LSG 
with a 1 TeV, in all flux models and both flavor scenarios 
with a 2-3 years of running. An interesting feature of the LSG (2TeV) entries in 
Tables~\ref{tab:table3} and~\ref{tab:table4} is that the down shower events may be enhanced 
enough compared to SM to distinguish between the two in WB, MB, and PR, and possibly SD 
too. The ratio of ratios in Tables~\ref{tab:table4} and~\ref{tab:table6} compares 
the LSG shower 
down/muon up ratios to the ones for SM. Table IV and VI also show the same 
for taus down/taus up. This diagnostic is especially sensitive to the difference between SM 
and LSG. It also shows us that this ratio of the ratios for showers and muons is 
almost the same in both of the flavor scenarios. The importance of the taus 
in differentiating between SM and LSG (1 TeV) is realized by looking at  the tagged 
tau events (Tables IV and VI). For tagged tau events, the difference between LSG (1 TeV) and 
the SM varies from an order of magnitude to two orders of magnitude for energy thresholds 
of 0.5-5 PeV. However, we caution the reader again that the statistics are low in this case. 
\textit{Basically the tau story can be summarized by saying that any upward tau event 
establishes 
(1) $\nu_{\tau}$'s presence in the neutrino flux incident on earth (2) exclusion of LSG 
with 1 TeV scale or any model of enhanced 
cross section of comparable size in the 1-10 PeV range.}  

The 5 PeV threshold results in Table V and VI show the same patterns as in the 0.5PeV 
tables. The distinction between SM and LSG (2 TeV) are now sharper in the ratios, but the 
statistics in some cases are low, so that one needs to have 5-10 years of data to draw 
strong conclusions that apply to all flux models. 

We conclude on the basis of our flux and event rate study that with a threshold of 
0.5 PeV, the shower and muon event ratios have sufficient events in all flux and lepton
flavor models to make clear distinctions between SM and LSG with a mass scale 2 TeV and below. 
Going above 2 TeV, one finds that whether distinctions can be made depende upon the 
flux model. The situation is not so clear. Because the requirements on $\nu_{\tau}$ 
identification are so stringent, only a few 
events to a fraction of an event will be expected, depending upon flux model, 
up or down event and LSG scale value. One point is perfectly clear: any upward tau event 
excludes LSG with a scale around 1TeV.

Given the intense
experimental activity in the field, we expect that data will yield many
insights in the coming decade when analysed with techniques like the ones
presented here.

\begin{center}
{\Large Acknowledgments}
\end{center}

We thank Pankaj Jain for many helpful comments and for use of his
propagation programs at the early stage of this work. Shahid Hussain thanks
M. H. Reno, J. Pumplin, and C. Hettlage for email exchange and discussions.
This work was supported in part by The Department of Energy under grant No.
DE-FG03-98ER41079. We used the computational facilities of the Kansas Center
for Advanced Scientific Computing for part of this work.

\end{document}